\documentclass[aps,showpacs,preprintnumbers,superscriptaddress,nofootinbib,twocolumn]{revtex4}

\usepackage[dvips]{graphicx}
\usepackage{bm,latexsym,amsmath,amssymb,amsfonts,mathrsfs}

\usepackage{amsmath}
\usepackage{bm}
\usepackage{multirow}
\usepackage{comment}

\usepackage{color}
\input{colordvi.tex}

\newcommand{\simgt}{\lower.5ex\hbox{$\; \buildrel > \over \sim \;$}}
\newcommand{\simlt}{\lower.5ex\hbox{$\; \buildrel < \over \sim \;$}}

\def\vgamma{{{\vec \gamma}}}

\begin{document}

\title{Observational Constraints on Kinetic Gravity Braiding from the Integrated Sachs-Wolfe Effect}

\author{Rampei~Kimura}
\affiliation{
Department of Physical Science, Hiroshima University,
Higashi-Hiroshima 739-8526,~Japan}

\author{Tsutomu~Kobayashi}
\affiliation{Hakubi Center, Kyoto University, Kyoto 606-8302, Japan}
\affiliation{Department of Physics, Kyoto University, Kyoto 606-8502, Japan}

\author{Kazuhiro~Yamamoto}
\affiliation{
Department of Physical Science, Hiroshima University,
Higashi-Hiroshima 739-8526,~Japan}


\begin{abstract}
The cross-correlation between the integrated Sachs-Wolfe (ISW) effect
and the large scale structure (LSS) is a powerful tool to
constrain dark energy and alternative theories of gravity. 
In this paper, we obtain observational constraints
on kinetic gravity braiding from
the ISW-LSS cross-correlation. 
We find that the late-time ISW effect 
in the kinetic gravity braiding model
anti-correlates with large scale structures in a wide range of parameters,
which clearly demonstrates how one can distinguish modified gravity theories 
from the $\Lambda$CDM model using the ISW effect. 
In addition to the analysis based on a concrete model,
we investigate a future prospect of the ISW-LSS cross-correlation 
by using a phenomenological parametrization of modified gravity models.
\end{abstract}

\maketitle


\def\bfp{{\bf p}}
\def\bfk{{\bf k}}
\def\bfx{{\bf x}}
\def\bfA{{\bf A}}
\def\bfv{{\bf v}}
\section{Introduction}
Modifying general relativity at long distances
is a possible way to explain the present accelerated expansion of the Universe
indicated by different
cosmological observations such as type Ia supernovae~\cite{Riess,Perlmutter}, 
the cosmic microwave background (CMB) anisotropies~\cite{Spergel,Komatsu}, 
the large scale structure (LSS) of galaxies~\cite{Percival,sloan}, 
and clusters of galaxies~\cite{Allen,Rapetti}.
A number of modified gravity models have been proposed so far,
including scalar-tensor theories,
$f(R)$ gravity~\cite{Carroll,Nojiri,Capozziello,HuSawicki,Starobinsky,
Tsujikawafr,Nojirib}, 
the Dvali-Gabadadze-Porrati (DGP) brane model~\cite{DGP1,DGP2},
and galileon gravity~\cite{GALMG}.

Recently, the galileon gravity model has attracted considerable attention,
which is constructed by introducing the scalar field with 
the self-interaction whose Lagrangian is invariant in the 
Minkowski space-time under the Galilean symmetry 
$\partial_{\mu}\phi \to \partial_{\mu}\phi+b_{\mu}$, 
which keeps equation of motion the second order 
differential equation with no ghostlike instabilities \cite{GALMG}. 
A remarkable property of the galileon model is the 
Vainshtein mechanism that the self-interaction terms induce 
the decoupling of the galileon field $\phi$ from gravity at 
small scale \cite{Vainshtein}. 
This allows the galileon theory to recover general relativity 
around a high density region, which ensures the consistency 
with the solar system experiments.
Furthermore, the galileon model can lead to the late-time 
accelerated expansion of the universe, which can be useful 
as an alternative to the simple dark energy model.

Cosmology of the galileon model has been extensively investigated 
by many authors \cite{CG,DDEF,GC,SAUGC,ELCP,CEGH,GGC,GGIR,CCGF,CSTOG,FeliceTsujikawa,Ginf,Kamada:2010qe,Kobayashi:2011pc,Kobayashi:2011nu,Gao:2011vs,geff,Qiu:2011cy, Easson:2011zy}.
Recently, Deffayet {\it et al.} \cite{GenGalileon} proposed
the most generalized scalar-tensor theory, 
including $f(R)$ theories, galileon theories 
and the kinetic gravity braiding \cite{Deffayet,Deffayet2,Kimura}.
In addition, several authors have investigated the observational constraints 
on various galileon theories from type Ia supernovae, the CMB shift parameter, 
baryon acoustic oscillations and the growth rate
of matter density perturbations \cite{MGALG,OCG,hirano1,hirano2}.

It was pointed out that the integrated Sachs-Wolfe effect can 
distinguish between the galileon model and the $\Lambda$CDM model, 
which was discussed in the scalar-tensor galileon model \cite{ELCP}. 
In contrast to the $\Lambda$CDM model, 
in which the gravitational potentials decay 
because of the accelerated expansion of the universe,
the amplitude of the gravitational potentials 
in the galileon model may increase 
due to perturbations of the galileon field.
In these two models, the integrated Sachs-Wolfe effect predicts 
opposite signs.  
Therefore, this provides us with a useful chance to test the 
galileon model using a cross-correlation between the matter 
(galaxy) distribution and the cosmic microwave background 
anisotropies \cite{CT1,BCT}.
In the present paper, we focus our investigation on 
the observational constraints from the
cross-correlation between the CMB anisotropies and the 
galaxy distributions.

Throughout the paper, we use units in which the speed of light and the Planck
constant are unity, $c=\hbar=1$, and $M_{\rm Pl}$ is the reduced Planck mass
related with Newton's gravitational constant by $M_{\rm Pl}=1/\sqrt{8\pi G}$. 
We follow the metric signature convention $(-,+,+,+)$.

\section{Kinetic gravity braiding model}
We start with the most general minimally coupled
scalar field theory with second-order field equations~\cite{Deffayet, Ginf},
whose action is of the form
\begin{eqnarray}
  S=\int d^4x \sqrt{-g}\left[\frac{M_{\mathrm{Pl}}^2}{2}R
  +K(\phi,X)-G(\phi,X)\square \phi+\mathcal{L}_{\mathrm{m}} \right],
\label{Lagrangian}
\end{eqnarray}
where $R$ is the Ricci scalar, $K(\phi,X)$ and $G(\phi,X)$ are 
arbitrary functions of the scalar field $\phi$ and
its kinetic term 
$X:=-g^{\mu\nu}\nabla_\mu\phi\nabla_\nu\phi /2$, 
$\square\phi=g^{\mu\nu}\nabla_\mu\nabla_\nu\phi$, 
and $\cal{L}_{\mathrm{m}}$ is the matter Lagrangian. 
Although $\phi$ is not directly coupled to $R$,
the interaction $G\square\phi$ causes ``braiding'' of the scalar field and the metric,
giving rise to non-trivial and interesting cosmology.
In this paper, we consider the shift symmetric case, $K=K(X)$ and $G=G(X)$,
for which an attractor solution is generically present~\cite{Deffayet, Kobayashi:2011nu},
leading to a late-time de Sitter expansion driven by
constant kinetic energy of $\phi$.
In particular, we focus on the following model~\cite{Kimura}:
\begin{eqnarray}
&&K(X)=-X,
\label{K}
\\
&&G(X)=M_{\mathrm{Pl}} \left(\frac{r_c^2}{M_{\mathrm{Pl}}^2}X \right)^n,
\label{G}
\end{eqnarray}
where $n$ and $r_c$ are the model parameters, though 
$r_c$ is required to be tuned as $r_c \sim H_0^{-1}$.
Throughout the paper
we consider for simplicity the cosmological evolution along the attractor,
assuming that the background solution converges to
the attractor at some sufficiently early time.

We work in the spatially flat Friedmann-Robertson-Walker Universe
and consider the metric perturbations in the conformal Newtonian gauge, 
\begin{eqnarray}
 ds^{2}=a^2(\eta)\biggl[-(1+2\Psi)d\eta^{2}
 +(1+2\Phi)(d\chi^2+\chi^2d\Omega_{(2)}^2)\biggr],
\end{eqnarray}
where the scale factor $a$ is normalized to unity at the present epoch,
$\eta$ is the conformal time, $\chi$ is the comoving distance 
and $d\Omega_{(2)}^2$ is the line element on a unit sphere.
The quasi-static approximation
is applicable for 
${\cal O}(k^2c_s^2/a^2) \gg {\cal O}(H^2)$,
which corresponds to $n\simlt 10$ for the wavenumber $k \simgt 0.01 h\rm{Mpc}^{-1}$.
Under this approximation, 
the modified Poisson equation may be written as
\begin{eqnarray}
\nabla^2 \Psi \simeq 4\pi G_{\rm eff} a^2\rho_m \delta,
\label{Pois1}
\end{eqnarray}
where $\rho_m$ is the energy density of matter, 
$\delta$ is the matter density contrast, and
$G_{\rm eff}$ is the effective gravitational coupling,
\begin{eqnarray}
 G_{\mathrm{eff}}
 =G\frac{2n+3n\Omega_{\rm m}(a)-\Omega_{\rm m}(a)}{\Omega_{\rm m}(a)[5n-\Omega_{\rm m}(a)]}.
  \label{geffomega}
\end{eqnarray}
The traceless part of the Einstein equations is given by
\begin{eqnarray}
  \Psi+\Phi=0.
\end{eqnarray}
The evolution equation for the matter overdensity $\delta$
is given by 
\begin{eqnarray}
\ddot{\delta}+2H\dot{\delta} \simeq \frac{\nabla^2}{a^2} \Psi.
\label{densityEq}
\end{eqnarray}

For large $n$, the quasi-static approximation no longer works.
In this case, one needs to solve the full perturbation equations.
The full perturbation equations and 
the validity of the quasi-static approximation are investigated in \cite{Kimura}.
For convenience, we define $U_k(\eta)$ as 
\begin{eqnarray}
-k^2\frac{\Psi_\bfk-\Phi_\bfk}{2}={3\over 2}\Omega_0H_0^2 U_k(\eta) \delta_\bfk(\eta_0),
\label{PoissonEq}
\end{eqnarray}
where $\Psi_\bfk$, $\Phi_\bfk$ and $\delta_\bfk$ are the metric perturbations
and the energy density contrast in Fourier space, respectively.
As long as the quasi-static approximation holds, 
$U_k(\eta)=(G_{\rm eff}/G)(D_1(\eta)/aD_1(\eta_0))$,
where $D_1(\eta)$ is the growth factor of the kinetic gravity braiding model
normalized as $D_1(\eta)=a$ at early stage of the evolution and $\eta_0$ is the present conformal time.

The authors of Ref.~\cite{Kimura} obtained the 
observational constraints on the kinetic gravity braiding model
with the functions~(\ref{K}) and~(\ref{G}) from 
type Ia supernovae, the WMAP
CMB anisotropy experiment, and the Sloan Digital Sky Survey (SDSS)
luminous red galaxy (LRG) samples. 
The result of Ref.~\cite{Kimura} indicates
that models with larger $n$ better fit the CMB distance observation,
though the constraints derived there are not so strong.
In this paper, we will improve the constraints significantly
by using the galaxy-CMB cross-correlation.

\section{ISW-Galaxy Cross-correlation}
The ISW effect contributes to the power spectrum of
the CMB anisotropy on large scales.
However, the direct detection of the 
ISW effect is very difficult due to the cosmic variance and 
the relatively small amplitude of the ISW effect
in the power spectrum. Nevertheless,
by cross-correlating the CMB anisotropies
and the galaxy number density fluctuations, one can
isolate the ISW effect, which will be a powerful probe of
dark energy and modified gravity models.
The ISW effect in the CMB anisotropy is given by
\begin{eqnarray}
{\Delta T(\vgamma)\over T}=\int_{\eta_d}^{\eta_0} d\eta [
\Psi'(\eta, \bfx)-\Phi'(\eta, \bfx)],
\end{eqnarray}
where $\eta_d$ is the decoupling time, which may be taken to be zero 
practically, and the prime denotes the differentiation 
with respect to $\eta$. 
The fluctuations in the angular distribution 
of galaxies is given by 
\begin{eqnarray}
{\Delta N_g(\vgamma)\over N}=
\int_0^{z_d} dz \delta_g(\eta,\bfx){\cal W}(z),
\label{NN}
\end{eqnarray}
where ${\cal W}(\chi)$ is the selection function
and $\delta_g$ is the
galaxy number density contrast,
which is related to the energy density contrast $\delta$
through the bias $b$ as $\delta_g=b(k,z)\delta$.
The cross-correlation of the galaxy fluctuations and the CMB anisotropies
is thus expressed as 
\begin{eqnarray}
\biggl\langle{\Delta T(\vgamma)\over T}{\Delta N_g(\vgamma')\over N}
\biggr\rangle
={1\over 4\pi}\sum_\ell (2\ell+1)C_\ell {\cal P}_\ell(\mu),
\label{crossc}
\end{eqnarray}
where
${\cal P}_\ell$ is the Legendre polynomial with
$\mu$ being the cosine of the angle between $\vgamma$ and 
$\vgamma'$,
\begin{eqnarray}
&&C_\ell={3\Omega_0H_0^2\over (\ell+1/2)^2}\int dz  {H(z){\cal W}(z)}
{D_1(z)\over D_1(z=0)}
\nonumber\\
&&~~~~~~\times {dU_k(\eta)\over dz }{b(z,k) P\left(k\right)}\biggr|_{k=(\ell+1/2)/\chi},
\label{cl}
\end{eqnarray}
and 
$P(k)$ is the matter power spectrum at the present time, $\eta=\eta_0$.
In deriving the above equation, we used the small angle approximation, $l \gg 1$.

\section{Comparison with observations}

\begin{figure*}[tb]
  \begin{center}
    \includegraphics[scale=1.0]{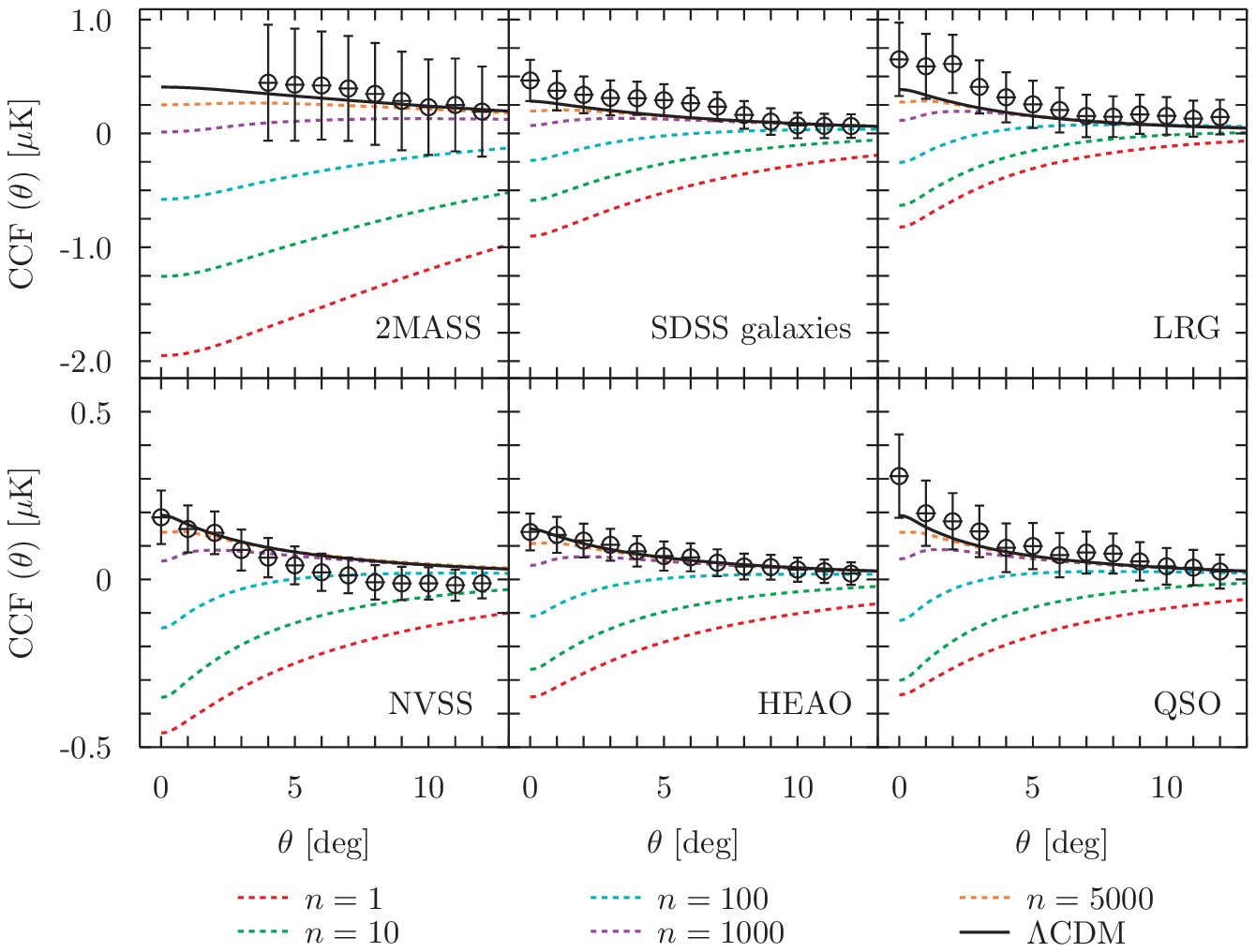}%
  \end{center}
\caption{The cross-correlation function theoretically calculated 
by using Eq.~(\ref{crossc}) and the data obtained in \cite{Giannantonio08a}.
Each curve shows the cross-correlation function of the $\Lambda$CDM model (solid curve), 
the KGB model (dashed curve) with $n=5000,~1000,~100,~10,$ and $1$, 
from the top to the bottom, respectively.
\label{fig_ccf} {}}
\end{figure*}

Figure~\ref{fig_ccf} shows the measured cross-correlation function 
for the six catalogues (2MASS, SDSS galaxies, LRG, NVSS, HEAO, and QSO)
available in Ref.~\cite{Giannantonio08a}. 
\footnote{
Most ISW detections have less than $3\sigma$ detection level
\cite{BC04,BC02,BC05,Giannantonio06,Giannantonio08a,fosalba03,cabre06,FrancisPeacock,Xia09,ho08,granett09,afshordi04,rassat07,nolta04,pietrobon06,Lopez10,Scranton,Padmanabhan05,Vielva06,McEwen08,McEwen07,Raccanelli08,Hernandez10,Corasaniti05,Gaztanaga06}.
In addition, the authors in Ref.~\cite{sawangwit10} claimed
the possibility of the absence of ISW signal 
extracted from the five-year WMAP data and the AAOmega LRG redshift survey. 
Although the detection of the cross-correlation between 
the cosmic microwave background and the large scale structure 
has been under debate (see, e.g., \cite{Sawangwit}),
we use the result of a combined analysis 
of the six catalogues in Ref.~\cite{Giannantonio08a}, 
which is the highest detection ($\sim 4.5\sigma$).
}
The solid and dashed curves show the theoretical predictions
for the $\Lambda$CDM model and the kinetic braiding model
with $n=1,~10,~100,~1000,$ and$~5000$.
We fixed the baryon density as $\Omega_b=0.0451$, 
the matter density as $\Omega_0h^2=0.1338$,
the hubble constant as $h=0.702$, the spectral index as $n_s=0.966$, 
and the amplitude of the density fluctuation as $\Delta_{\cal R}^2=2.42\times 10^{-9}$ 
at $k_0=0.002~ {\rm Mpc^{-1}}$~\cite{Komatsu}.
The galaxy bias of each catalog is found in Ref.~\cite{Giannantonio08a} and 
the values are $b=1.4,~1.0,~1.8,~1.5,~1.06$ and $2.3$ for 
2MASS, SDSS galaxies, SDSS LRG, NVSS, HEAO, and QSO, respectively.
For the theoretical modeling of the power spectrum, 
we used the linear power spectrum with 
the transfer function $T(k)$ of Eisenstein and Hu~\cite{EHu}
and obtained $\Psi$, $\Phi$, and $\delta$ by solving the full perturbation
equations. %
As we can see from Fig.~\ref{fig_ccf}, 
the cross-correlation function for the kinetic gravity braiding model
with larger $n$ is closer to that for the $\Lambda$CDM model,
and they are practically indistinguishable for $n>{\cal O}(1000)$.
This is because the perturbation of the kinetic term $\delta X$ become zero 
in the limit of $n \to \infty$~\cite{Kimura}. Another important feature
found in Fig.~\ref{fig_ccf} is the anti-correlation 
for the kinetic gravity braiding model with small $n$.
This is caused by the enhancement of the effective gravitational coupling $G_{\rm eff}$
and the consequent growth of the gravitational potential,
leading to the opposite sign of the function $dU_k(z)/dz$
compared to the $\Lambda$CDM model.
Since the measurements of the cross-correlation function 
in each catalog shows a positive correlation,
kinetic gravity braiding with small $n$ will be inconsistent with observations.

In our analysis, the total chi squared is given by 
\begin{eqnarray}
  \chi^2_{\rm total}=\sum_{i,j}
  (c_i^{\rm obs}-c_i^{\rm theo}) C_{ij}^{-1} (c_j^{\rm obs}-c_j^{\rm theo}),
\end{eqnarray}
where $c_i^{\rm obs}$ is the cross-correlation function 
obtained from observations, $c_i^{\rm theo}$ is 
the cross-correlation function theoretically predicted 
from Eq.~(\ref{crossc}), and $C_{ij}^{-1}$ is the inverse of 
the covariance matrix obtained from \cite{Giannantonio08a}.
For the WMAP cosmological parameters \cite{Komatsu}, we obtained 
the lower bound
\begin{eqnarray}
n> 4.2 \times 10^3
\quad
(95\% {\rm ~C.L.}).
\label{bound}
\end{eqnarray}
As we expected from Fig.~\ref{fig_ccf}, 
the kinetic gravity braiding model with small $n$ 
is obviously ruled out, while the model
with large $n$ is favored by observations. 
This is because the ISW effect anti-correlates
with the LSS for $n < {\cal O}(1000)$,
and the sign of the cross-correlation function 
determines the lower bound (\ref{bound}).
Note that the galaxy bias in Ref. \cite{Giannantonio08a} 
is obtained from the two-point correlation functions 
of each catalog assuming the $\Lambda$CDM model.
In our case, the matter power spectrum for $n \simgt 1000$ 
and $0.001~h{\rm Mpc}^{-1} \simlt k \simlt 0.1~h{\rm Mpc}^{-1}$, 
which is the dominant wavenumber in Eq.~(\ref{cl}),
is almost the same as those in the $\Lambda$CDM model.
Therefore, this is not problematic, 
and we also confirmed that using the galaxy bias of the $\Lambda$CDM model 
does not significantly change the results of chi-squared analysis.
The lower bound (\ref{bound}) changes within only $10$ percent even when we assume 
the bias as free parameters. This is because the constraint on 
the model parameter $n$ is determined by the sign of the 
cross-correlation function, which is almost independent
of nature of bias. 

\section{Parametrized model}
%

In the previous section, we have demonstrated
that the ISW-galaxy cross-correlation strongly constrains
kinetic gravity braiding. The key to this result was the
{\it time-evolution} of the gravitational potential controlled 
by the effective gravitational coupling $G_{\rm eff}$
in the density perturbation equation.
In this section, we employ a phenomenological parametrization 
of the {\it time-evolution} of $G_{\rm eff}$
rather than $G_{\rm eff}$ derived from a concrete model,
and investigate how the parameters are constrained by tomographic 
observations. This approach is useful to estimate a general 
prospect of constraints on generalized models (e.g., \cite{geff}).

Provided that the quasi-static approximation is valid,
the evolution of overdensities is described by the equation
of the form~(\ref{densityEq}) with~(\ref{Pois1})
in {\em all} the scalar-tensor theories with second-order field equations~\cite{geff}
including kinetic gravity braiding.
We parametrize the evolution of $G_{\rm eff}$ as follows:
\begin{eqnarray}
\frac{G_{\rm eff}}{G}=1+g_1 a^{g_2},
\label{geff_p}
\end{eqnarray}
where $g_1$ and $g_2$ are the free parameters.
This reproduces the effective gravitational coupling e.g., of 
the kinetic gravity braiding model considered in this paper for small $n$.
In particular, the effective gravitational coupling with $g_1=1$ and $g_2=4$ yields
that of the kinetic gravity braiding model for $n=1$ within a few percent precision.
We consider the model whose sound speed for the scalar field does not approach zero.
This means the quasi-static approximation is applicable for the dominant wavenumbers 
in Eq.~(\ref{cl}).
Then, 
the evolution equation for the matter overdensity and 
the Poisson equation are of the form Eqs.~(\ref{densityEq}) and (\ref{PoissonEq}).
In the case of a minimally coupled scalar field model, 
the traceless part of the Einstein equations gives $\Psi+\Phi=0$.

\begin{figure}[tb]
  \begin{center}
    \includegraphics[keepaspectratio=true,height=67mm]{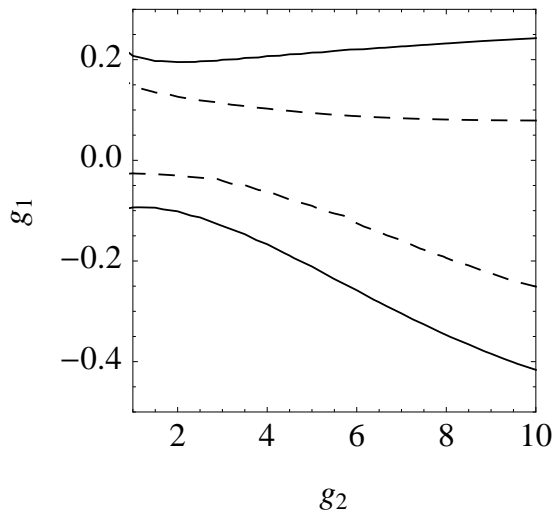}
  \end{center}
\caption{Contour of $\Delta\chi^2$ on the $g_1-g_2$ plane. 
The dashed curve and solid curve are the $1\sigma$ and 
2$\sigma$ confidence contour-levels, respectively.
\label{fig_contour} {}}
\end{figure}

In this section, we present the observational constraints 
on the above parametrized model and perform the Fisher matrix analysis. 
Figure~\ref{fig_contour} shows the contour of $\Delta\chi^2$ on the $g_1-g_2$ plane. 
We
assumed that the background expansion history is
the same as that of the standard $\Lambda$CDM model,
and
used the cosmological parameters from the WMAP7 data~\cite{Komatsu}.
The background history for $n>100$ is almost identical to
the LCDM model. As seen from the section IV, for $n<100$ the ISW-LSS cross-correlation is
negative, which is clearly incompatible with observational data, and 
the different background evolution does not change the sign of the correlation
function and hence is not important.
For each parameter, we assume that the galaxy bias is determined by 
the amplitude of the power spectrum of the matter distribution,
$b^{(i)}=b^{(\Lambda{\rm CDM)}}D_1(z_*^{(i)})/D_1^{(\Lambda{\rm CDM)}}(z_*^{(i)})$, 
where $b^{(\Lambda{\rm CDM)}}$ is the bias used in the previous section, 
$D_1^{(\Lambda{\rm CDM)}}$ is the growth factor of the $\Lambda$CDM model, 
and $z_*^{(i)}$ is the mean redshift of the $i$-th catalog.
Our result implies that the deviation of the effective gravitational coupling
from Newton's constant must be very small, 
as expected in~\cite{zhao10}.
Since $g_1 \to 0$ or $g_2 \to \infty$ corresponds to the $\Lambda$CDM model for $a<1$
the constraint on $g_2$ can not be determined as long as the $\Lambda$CDM model
is favored by observations.

\begin{figure}[t]
  \begin{center}
    \includegraphics[keepaspectratio=true,height=67mm]{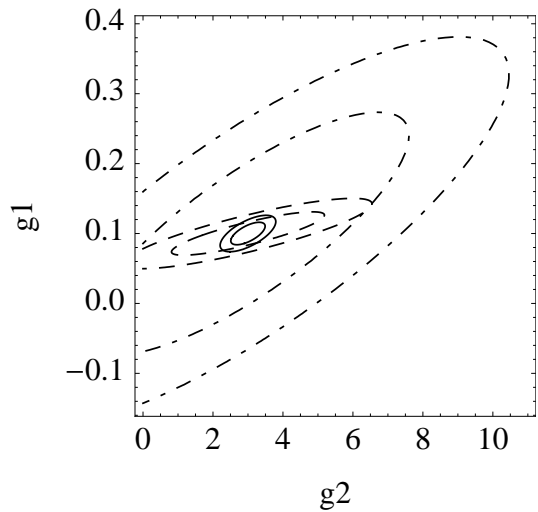}
  \end{center}
\caption{The 1-sigma and 2-sigma contours 
in the $g_1-g_2$ plane for the future survey. 
We assumed ${\cal A}=1500~{\rm deg}^2$ + 4 bins (dot-dashed curve), ${\cal A}=20000~{\rm deg}^2$ (dashed curve),
${\cal A}=20000~{\rm deg}^2$ + 4 bins (solid curve).
The target parameter is $g_1=0.1$ and $g_2=3$.
\label{fig_fisher} {}}
\end{figure}

Next, we give a forecast for the accuracy of constraining $g_1$ and $g_2$ from future galaxy surveys.
In the Fisher matrix analysis \cite{tegmark97}, we consider the following redshift 
distribution of the galaxy sample per unit solid angle,
\begin{eqnarray}
  \frac{dN}{dz}=\frac{N_g \beta}{z_0^{\alpha+1}\Gamma((\alpha+1)/\beta)}z^{\alpha}
  \exp\left[-\left(\frac{z}{z_0}\right)^{\beta}\right],
\end{eqnarray}
where $\alpha$, $\beta$, $z_0$ are the parameters, 
and $N_g=\int dN/dz$. The mean redshift is determined by
\begin{eqnarray}
  z_m=\frac{1}{N_g}\int dz z \frac{dN}{dz}
=\frac{z_0\Gamma((\alpha+2)/\beta)}{\Gamma((\alpha+1)/\beta)}.
\end{eqnarray}
We used $\alpha=0.5$, 
$\beta=3$, $N_g=35~{\rm arcmin}^{-2}$ and $z_m=0.9$.
We divide the galaxy sample 
into four redshift bins, $0.05<z<0.6 z_m$,~$0.6 z_m<z<z_m$,
~$z_m<z<1.4 z_m$, and $1.4 z_m<z<2.5$ \cite{yphn}, 
where the redshift distributions of 
i-th bin ($z_{i-1} \leq z \leq z_i$) are given by \cite{hu04}
\begin{eqnarray}
  W_i(z)=\frac{A_i}{2}\frac{dN}{dz}\left[{\rm erfc}
    \left(\frac{z_{i-1}-z}{\sqrt{2}\sigma(z)}\right)-{\rm erfc}
    \left(\frac{z_{i}-z}{\sqrt{2}\sigma(z)}\right)\right],
\nonumber\\
\end{eqnarray}
where erfc is the complementary error function,
$A_i$ is determined by the normalization constant,
and we adopted $\sigma(z)=0.03(1+z)$.
The Fisher matrix is given by \cite{douspis08,pogosian05,garriga04}
\begin{eqnarray}
  F^{ij}=f_{\rm sky}\sum_l (2l+1)\frac{\partial C_l^{\rm ISW-G}}{\partial \Theta_i}
  {\rm cov}^{-1}(l)\frac{\partial C_l^{\rm ISW-G}}{\partial \Theta_j},
\end{eqnarray}
where $f_{\rm sky}$ is the fraction of sky common to both the CMB and the galaxy survey maps, 
$C_l^{\rm ISW-G}$ is the two-point angular cross-correlation defined by Eq.~(\ref{crossc}),
$\Theta_i$ is the model parameter $g_1$ and $g_2$, and
\begin{eqnarray}
  {\rm cov}(l)=(C_l^{\rm ISW-G})^2+(C_l^{\rm ISW}+N_l^{\rm CMB})(C_l^{\rm G}+N_l^{\rm g}).
\end{eqnarray}
Figure~\ref{fig_fisher} shows the 1-sigma (dashed curve) 
and 2-sigma (solid curve) confidence contours in the $g_1-g_2$ plane,
assuming future survey such as the HSC-like weak lensing survey 
(${\cal A}=1500~{\rm deg^2}$) and the LSST-like survey (${\cal A}=20000~{\rm deg^2}$). 
The target model is the parametrized model given by Eq.~(\ref{geff_p}) 
with $g_1=0.1$ and $g_2=3$. We used the WMAP cosmological parameters 
for $\Omega_0 h^2$, $\Omega_b h^2$ and $\Delta_{\cal R}$ and
assumed that the bias and the initial amplitude of the
fluctuation is determined by individual observations such as CMB power spectra and 
matter power spectra. 
Then we consider only 2 parameters $g_1$ and $g_2$ 
in the Fisher matrix analysis.
As one can see from Fig.~\ref{fig_fisher}, the tomography, 
by dividing the galaxy sample into several redshift bins, can significantly 
improve constraints on the parameter $g_2$.
Therefore, the ISW-LSS cross-correlation 
with tomography can dramatically improve the determination
of the evolution of the effective gravitational coupling as well as 
its amplitude.

\section{Conclusion}
In this paper, we have obtained observational constrains 
on the kinetic gravity braiding model
from the cross-correlation function between 
the ISW effect and the large scale structure. 
We have found that the cross-correlation function of 
the kinetic gravity braiding model with small $n$ 
is negative, and therefore such a model is not 
favored by observations. Constraints 
from ISW-LSS cross-correlation is stronger than
those obtained from SNIa and the CMB shift parameter~\cite{Kimura}. 
Thus, the covariant galileon model, $n=1$,
is ruled out by the cross-correlation between the ISW effect 
and the large scale structure.

In deriving the constraints,
we have clarified that the evolution of the effective
gravitational coupling is crucial for the behavior of the ISW effect.
Motivated by this fact,
we also considered the model with a
phenomenological parametrization of $G_{\rm eff}$ that can 
reproduce the effective gravitational coupling of the covariant 
galileon model. We have confirmed that the deviation from 
the cosmological constant must be very small: 
the 1-sigma (2-sigma) error in determining $g_1$ is 
$\Delta g_1 \sim 0.15~(0.3)$ when $g_2\sim 2$.
We have performed the Fisher matrix analysis,
assuming a HSC-like survey and a LSST-like survey, and found that the
ISW-LSS cross-correlation
with the sample divided into multiple redshift bins can improve significantly
the constraint on $g_2$. This tells us that the ISW-LSS
cross-correlation
obtained from the future survey would be a powerful probe 
for long distance modification of general relativity.

\vspace{0.2cm}
{\it Acknowledgment~~}
We thank K. Koyama for useful comments. 
This work was supported by Japan Society for Promotion
of Science (JSPS) Grants-in-Aid
for Scientific Research (No.~21540270 and No.~21244033)
and JSPS Grant-in-Aid for Research Activity Start-up
No. 22840011.
K.Y. thanks R. G. Crittenden for useful conversation on the
topic of the present paper.
This work was also supported by JSPS Core-to-Core Program 
``International Research Network for Dark Energy''.

\end{document}